\documentclass[fleqn,12pt,twoside,titlepage]{article}

\usepackage{graphicx}
\usepackage[figuresright]{rotating}

\newcommand{\AmS}{{\protect\the\textfont2
  A\kern-.1667em\lower.5ex\hbox{M}\kern-.125emS}}

\title{The Kr2Det project:
Search for mass-3 state contribution $|{U_{e3}}|^2$ to the electron neutrino 
using a one reactor - two detector oscillation experiment at 
Krasnoyarsk underground site.}

\author{V. Martemyanov\thanks{vpmar@dnuc.polyn.kiae.su}, 
L. Mikaelyan\thanks{lmikael@polyn.kiae.su, lmikael@ostrov.net},
	  V. Sinev\thanks{sinev@polyn.kiae.su}, \\
V. Kopeikin\thanks{kopeykin@polyn.kiae.su}, 
Yu. Kozlov\thanks{kozlov@dnuc.polyn.kiae.su}\\
\\
Russian Research Center "Kurchatov Institute", \\
Kurchatov Square 1, Moscow, 123182, Russia}

\date{}

\begin{document}

\maketitle

\tableofcontents

\begin{abstract}
The main physical goal of the project is to search with reactor antineutrinos 
for small mixing angle oscillations in the atmospheric mass parameter 
region around ${\Delta}m^{2}_{atm} \sim 2.5 \times 10^{-3}$ eV$^2$ in order to find the element 
$U_{e3}$ of the neutrino mixing matrix or to set a new more stringent constraint 
($U_{e3}$ is the contribution of mass-3 state to the electron neutrino flavor state). 
To achieve this we propose a ``one reactor - two detector" experiment: two 
identical antineutrino spectrometers with $\sim$50 ton liquid scintillator targets 
located at $\sim$100 m and $\sim$1000 m from the Krasnoyarsk underground reactor 
($\sim$600 mwe). In no-oscillation case ratio of measured positron spectra of 
the $\bar{{\nu}_e}  + p \rightarrow e^{+} + n$ reaction is energy independent. Deviation 
from a constant value of this ratio is the oscillation signature. In this scheme results 
do not depend on the exact knowledge of the reactor power, $\bar{{\nu}_e}$ spectra, 
burn up effects, target volumes and, which is important, the backgrounds 
can periodically be measured during reactor OFF periods. In this letter we 
present the Krasnoyarsk reactor site, give a schematic description of the 
detectors, calculate the neutrino detection rates and estimate the 
backgrounds. We also outline the detector monitoring and calibration 
procedures, which are of a key importance. We hope that systematic 
uncertainties will not accede 0.5\% and the sensitivity $U^{2}_{e3} \approx 4\times 10^{-3}$ 
(at ${\Delta}m^{2} = 2.5 \times 10^{-3}$ eV$^2$) can be achieved.
\end{abstract}

\section{Introduction}

The Super-Kamiokande studies of atmospheric neutrinos [1] found 
intensive (${\sin}^{2}2{\theta}_{atm} > 0.9$) ${\nu}_{\mu}\rightarrow {\nu}_{x}$ 
oscillations and have confined the mass parameter to the interval 
$1.4\times 10^{-3} < {\Delta}m^{2}_{atm} < 4.2 \times 10^{-3}$ eV$^2$ with 
${\Delta}m^{2}_{atm}= 2.5\times 10^{-3}$ eV$^2$ as the most probable value. 
The ${\nu}_{\mu}\rightarrow {\nu}_{\tau}$ has been found to be the dominant 
channel of the atmospheric neutrino oscillations, while much place is left 
also for the ${\nu}_{\mu}\rightarrow {\nu}_{{\mu},{\tau}}$ transitions. 

The $\sim$ 1 km baseline reactor experiment CHOOZ [2] searched for electron antineutrino 
disappearance in the atmospheric mass parameters region. No oscillation have been found 
(Fig.1, curve ``CHOOZ"):
\begin{eqnarray}
\sin^{2}2{\theta}_{CHOOZ} \le 0.14 \quad (90\% \ {\rm CL \ at} \ {\Delta}m^2 = 2.5\times 10^{-3} {\rm eV}^2),\nonumber\\
\sin^{2}2{\theta}_{CHOOZ} \le 0.19 \quad ({\rm at} \ {\Delta}m^2 = 2.0\times 10^{-3} {\rm eV}^2)
\end{eqnarray}

The reactor neutrino mixing parameter $\sin^{2}2{\theta}$ in atmospheric mass parameters 
region plays an important role in the neutrino oscillation physics. In three active neutrino 
mixing scheme with normal neutrino mass hierarchy it is expressed trough the contribution of 
the mass-3 eigenstate to the electron neutrino flavor state $U_{e3} = \sin{\theta}_{13}$,:

\begin{equation}
\sin^{2}2{\theta}_{CHOOZ} = \sin^{2}2{\theta}_{13}=4|U_{e3}|^2 (1 - |U_{e3}|^2)
\end{equation} 
We mention also that with nonzero value of $U_{e3}$ in the lepton sector the CP violation 
effects can exist.
 
The negative results of the CHOOZ experiment impose important constraint: 

\begin{equation}
\sin^{2}2{\theta}_{13} \le 0.14, \ |U_{e3}|^2 \le 3.6\times 10^{-2} \ ({\rm \ at} \ {\Delta}m^2 = 2.5\times 10^{-3} {\rm eV}^2)
\end{equation} 
	
The quantity $\sin^{2}2{\theta}_{13}$ can be hundreds and thousands times smaller than 
present CHOOZ limits. In this case necessary sensitivity can, in distant future and in 
several steps, be achieved at Neutrino Factories in experiments using hundred and thousand 
kt detectors located at a few thousand km from the accelerator neutrino source (For review 
see Ref. [3]). 
 
The first step can, however, be done sooner (and cheaper) at reactors as has been discussed 
since 1999 y. [4,5]. Recently an idea to search for $U_{e3}$ at reactors in Japan was 
published [6]. To do this first step is still more important because no physical reason is 
known why $\sin^{2}2{\theta}_{13}$ should be very small. It may quite happen that this 
quantity is only several times smaller than present upper limits (1). 
 
The main physical goals of the reactor experiment considered here are:
\begin{itemize}
\item To obtain new information on the electron neutrino mass composition ($U_{e3}$),
\item To provide normalization for future experiments at accelerators,
\item To achieve better understanding of the role ${\nu}_e$ can have in the atmospheric neutrino phenomena.
\end{itemize}
The main practical goal is to decrease, relative to CHOOZ, statistic and systematic 
uncertainties as much as possible.
 
Analysis of all available solar neutrino data (Ref [7]) confirms the LMA MSW as the most 
probable solution with the best fit value of the solar neutrino mass parameter 
${\Delta}m^2_{sol} \approx 6\times 10^{-5}$ eV$^2$. We assume therefore that 
\begin{equation}
{\Delta}m^2_{sol} \ll {\Delta}m^{2}_{atm} = m^2_{3} - m^2_{2} \approx m^2_{3} - m^2_{1}
\end{equation}
and use in this paper two mode expression for the reactor antineutrino survival probability 
$P(\bar{{\nu}_e} \rightarrow \bar{{\nu}_e})$:
\begin{equation}
P(\bar{{\nu}_e} \rightarrow \bar{{\nu}_e}) = 1 - \sin^{2}2{\theta}\cdot \sin^{2}\left(\frac{1.27{\Delta}m^{2}L}{E}\right)
\end{equation}
where $L$(m) is the reactor - detector distance and $E$ (MeV) is the antineutrino energy. There is 
however some probability that ${\Delta}m^2_{sol}$ is not so small as assumed above. In this case 
${\Delta}m^2_{sol}/{\Delta}m^2_{atm}$ cannot be neglected and somewhat more complicated expressions 
for $P(\bar{{\nu}_e} \rightarrow \bar{{\nu}_e})$ should be used as discussed in Refs [6, 8].

Reactor antineutrinos have a continues energy spectrum and are detected via the inverse beta-decay 
reaction
\begin{equation}
\bar{{\nu}_e}  + p \rightarrow e^{+} + n
\end{equation}
The visible positron energy $E_e$ is related to the $\bar{{\nu}_e}$ energy as
\begin{equation}
E_e = E - 1.80 + E_{annih} \approx E - 0.8
\end{equation}
Typical positron energy spectrum is shown in Fig. 2
 
\section{One reactor - two detector scheme}

Two identical liquid scintillation spectrometers are stationed at distances $L_{far}\approx$ 1000 m 
(far position) and $L_{near}\approx$ 115 m from the underground Krasnoyarsk reactor. (Fig. 3) 
The overburden at Krasnoyarsk is $\sim$ 600 m.w.e., which is twice as much as in the CHOOZ 
experiment. (At short distances form the reactor the one reactor - 2 detector approach was first 
probed at Rovno [9] and later successfully used at Bugey [10])
 
Two types of analysis can be used.
Analysis I is based on comparison of the shapes of positron spectra $S(E_e)_{far}$ and $S(E_e)_{near}$ 
measured simultaneously in two detectors. In no oscillation case the ratio 
$S(E_e)_{far}/S(E_e)_{near}$ is energy independent. Small deviations from the constant value of this 
ratio 
\begin{equation}
X_{shape}= C\cdot \frac{1 - \sin^{2}2{\theta}\cdot \sin^{2}\left(\frac{1.27{\Delta}m^{2}L_{far}}{E}\right)}{1 - \sin^{2}2{\theta}\cdot \sin^{2}\left(\frac{1.27{\Delta}m^{2}L_{near}}{E}\right)}
\end{equation} 
are searched for oscillation parameters. 
 
In the one reactor - two detector scheme
\begin{itemize}
\item Results of the Analysis I do not depend on the exact knowledge of the reactor power, absolute 
$\bar{{\nu}_e}$ flux and energy spectrum, burn up effects, absolute values of hydrogen atom concentrations, 
detection efficiencies, target volumes and reactor - detector distances. 
\item At Krasnoyrsk the detector backgrounds can be measured during reactor OFF periods, which 
periodically follow 50 daylong reactor ON periods.  
\end{itemize}
Calculated ratios $S(E_e)_{far}/S(E_e)_{near}$ for a set of oscillation parameters are shown in 
Fig. 4.

Analysis II is based on the ratio of the total number of neutrinos $N_{far}, N_{far}$ detected 
at two distances:
$$
X_{rate}(\sin^{2}2{\theta},{\Delta}m^{2})= \left(\frac{L_{far}}{L_{near}}\right)^2\cdot \left(\frac{V_{near}}{V_{far}}\right)
\cdot\left(\frac{{\epsilon}_{near}}{{\epsilon}_{far}}\right)\cdot \left(\frac{N_{far}}{N_{near}}\right) \eqno (8')  
$$
$V_{far}, V_{near},{\epsilon}_{far},{\epsilon}_{near}$ are the target volumes and neutrino detection 
efficiencies.
In no oscillation case $X_{rate}$ = 1.
 
Analysis II is also independent of the exact knowledge of the reactor neutrino flux and energy 
spectrum. The absolute values of detection efficiencies are practically canceled, only their small 
difference is to be considered here while the ratios $(L_{far}/L_{near})^2$ and
$(V_{near}/V_{far})$ should be known accurately. 
 
\section{Detectors}

A miniature version of the KamLAND [11] and BOREXINO [12] and a scaled up version of the CHOOZ 
three - concentric zone detector design is chosen for the constraction of the spectrometers (Fig. 5). 
At this stage we consider 4.7 m diameter liquid scintillator target, enclosed in transparent 
spherical balloon. The target is viewed by $\sim$800 8-inch EMI-9350 (9350 - 9356) photomultipliers 
trough $\sim$90 cm layer of mineral oil of the zone-2 of the detector. The PMT's of this type have 
successfully been used in the CHOOZ experiment and are used now in the BOREXINO and SNO 
detectors [13]. A 20\% light collection and 150 - 200 photoelectron signal is expected for 1 MeV 
positron energy deposition. The PMT's are mounted on the stainless steel screen, which separates 
external zone-3 from the central zones of the detector. The $\sim$75 cm thick zone-3 is filled with 
mineral oil (or liquid scintillator) and serves as active (muon) and passive shielding from the 
external radioactivity. 

\section{Detector calibrations and monitoring; systematic uncertainties}

The ratio of measured positron spectra $S(E_e)_{far}/S(E_e)_{near}$ (Eq. 8) can be slightly distorted 
because of relative difference in response functions of the two ``identical" spectrometers. 

The goal of calibration procedures we consider is to measure this difference and introduce 
necessary corrections. This can be done by a combination of different methods. First we consider 
periodic control of the energy scales in many points using ${\gamma}$-sources shown by arrows in 
Fig. 2. A useful continuous monitoring of the scales at 2.23 MeV can provide neutrons produced 
by trough going muons and captured by the target protons during veto time. 

The second method uses small spontaneous fission $^{252}$Cf or $^{238}$U sources periodically 
placed in the detectors. These sources generates continuous energy spectrum due to prompt 
fission gammas and neutron recoils (the dashed line in Fig.2.). Deviation from unity of the 
measured spectra can be used to calculate relevant corrections.

We hope that systematic uncertainty due to detector spectrometric difference essential for 
Analysis I can be controlled down to 0.5\%. 

In Analysis II the systematic uncertainty in the quantity 
$(L_{far}/L_{
near})^2\cdot (V_{near}/V_{far})\cdot ({\epsilon}_{near}/{\epsilon}_{far})$
in Eqs (8`) can hopefully be kept within 0.8\%.

\section{Scintillator}

The final choice of the scintillator has not been made so far. 
We hope for progress in manufacturing Gd ($\sim$0.9 g/liter) loaded scintillators to improve 
the response to neutrons and suppress accidentals, which originate from U/Th gammas coming 
from surrounding rock. The Palo Verde Gd-scintillator showed better stability than the 
scintillator used in CHOOZ. The LENS project considers scintillators with rare earth contents 
as high as $\sim$50 g/liter.

Currently we consider no-Gd scintillator based on the mixture of isoparaffin or mineral oil 
and pseudicumene ($\sim$20\%) with $\sim$2g/liter PPO as primary fluor. This scintillator 
has C/H ratio 1.85, density 0.85 kg/liter and $0.785\times 10^{29} $ H atoms per ton.

\section{Neutrino detection rates and backgrounds}

The neutrino events satisfy the following requirements: (i) a time window on the delay between  
$e^{+}$ and neutron signals 2$-$600 ${\mu}$s, (ii) energy window for the neutron candidate 
1.7$-$3.1 MeV and for $e^{+}$ 1.2$-$8.0 MeV, (iii) distance between $e^{+}$ and neutron less 
than 100 cm. At this stage no pulse shape analysis to reject proton recoils is planned. 

Under these assumptions neutrino detection efficiency of 75\% was found and neutrino detection 
rate $N(e^{+},n)$ = 55/day calculated for the far detector. 

The time correlated background 0.1 per day per one target ton was found by extrapolation of 
the value 0.25/per day per target ton measured at CHOOZ:
\begin{equation} 
{\rm CHOOZ\ (300\ mwe),\ 0.25/day}\cdot{\rm ton} \rightarrow {\rm Kr2Det\ (600\ mwe),\ 0.1/day}\cdot{\rm ton}
\end{equation}

The accidental coincidences come from the internal radioactivity of detector materials and U 
and Th contained in the surrounding rock. The internal component of the background was estimated 
to be less 0.3/day, which is an order of magnitude smaller than the rate of the correlated 
background (see hep-ph/0109277). In contrast to the KamLAND and Borexino experiments three 
orders higher concentrations of U, Th, K and Rn can be tolerated in the liquids used in the 
Kr2Det case. 

First estimations of accidentals coming from the radioactivity of the rock showed however that 
external passive shielding of the detector should be increased in case scintillator without 
Gd is used as the neutrino target.
 
Calculated neutrino detection rates $N(e^{+},n)$ and backgrounds for scintillator with no Gd 
are summarized in the Table.
 
\begin{table}[htb]
\caption{}
\label{table:1}
\vspace{10pt}
\begin{tabular}{c|c|c|c|c|c|c}
\hline
Parameter  & Distance, & Target, & $N(e^{+},n)$, & $N(e^{+},n)$, & \multicolumn{2}{c}{Backgr., day$^{-1}$} \\
\cline{6-7}
           &     m     & mass, ton & day$^{-1}$    & year$^{-1*}$     & correl. & accid.$^{**}$ \\
\hline
Far detector & 1000 & 46 & 55 & $16.5\cdot 10^{3}$ & 5 & $\sim$0.3  \\
Near detector & 115 & 46 & 4200 & $12.5\cdot 10^{5}$ & 5 & $\sim$0.3  \\
\hline
\end{tabular}\\[2pt]
{\small $^{*}$ 300 days/year at full power.}\\
{\small $^{**}$ due to internal radioactivity of the detector materials only.}
\end{table}

\section{Expected results and Conclusions}

Expected 90\% CL constraints on the oscillation parameters (Fig. 1, curves K2Det) are obtained for 
40000 detected $\bar{{\nu}_e}$ in the far detector (750 days of full power). The systematic 
uncertainties ${\sigma}_{shape}$= 0.5\% in the Analysis I (``shape") and ${\sigma}_{rate}$= 0.8\% 
in the Analysis II (``rate") have been assumed. The ``shape" analysis is somewhat more sensitive 
and can shift (at ${\Delta}m^2=2.5\times 10^{-3}$ eV$^2$ the $\sin^22{\theta}$ upper limit from 0.14 
(CHOOZ) to 0.017. 

The one reactor - two detector approach fully eliminates uncertainties associated with the reactor 
neutrino source inherent to the absolute method used at CHOOZ.

Small relative difference in conceptually identical detector properties can be minimized through 
calibration and monitoring procedures.

The detector backgrounds can be measured during reactor OFF periods, which periodically follow 
50 daylong reactor ON periods.

Good signal to background ratio can be achieved due to sufficiently deep underground position of the 
detectors. 

High statistics can be accumulated in reasonably short time period using detectors with $\sim$45 ton 
targets, which are relatively small if compared to modern neutrino detectors.

Neutrino community has accumulated positive experience in building and running 3 concentric zone 
detectors similar to the Kr2Det detectors.

We conclude that proposed study is feasible and that new important information on the electron 
neutrino internal structure ($\sin^22{\theta}_{13}$) can be obtained.

\section*{Acknowledgments }

We are grateful to Yu. Kamyshkov and A. Piepke for fruitful discussions. This work is supported by 
RFBR grant Numbers 00-02-16035 and 00-15-98708.

\appendix

\section{Neutrino site at Krasnoyarsk}
The reactor belongs to the Federal State-Owned Unitary Enterprise MINING \& CHEMICAL COMBINE (MCC) 
53, Lenin Str., Zheleznogorsk, Krasnoyarsk Territory, RUSSIA, 660972.

The Krasnoyarsk neutrino laboratory is built in the MCC underground territory.

There are two places to install the detectors. One of them at $\sim$115 m from the reactor is 10 m 
high $15\times15$ m square room. The other is a 125 m long, 11.5 high and 15 m wide corridor at 
$\sim$1000 m from the reactor. More information on neutrino at Krasnoyarsk can soon be found at 
$http://www.lngs.infn.it/site/exppro/panagic/section\_indexes/ \\
frame\_particles.html$ (click 
``Laboratories and experiments", then ``Underground and underwater laboratories" and go to 
``Krasnoyarsk neutrino laboratory").

Zheleznogorsk is located at about 70km from Krasnoyarsk on the bank of the Yenisei River. 
Zheleznogorsk is a very nice and clean town built in direct neighborhood to the Siberian taiga, 
rich of birds and animals. There is a beautiful large lake in the center of the town. Picturesque 
hills surround the town center. A musicale theatre, hotel, rest home, restaurants, a lot of shops 
are in Zheleznogorsk.. The weather is comfortable; the number of sunny days is the same as in resort 
Sochi (at the Black Sea). Winter is cold but not so much compared with Moscow, air is dry. The summer 
and autumn are warmer and sunnier than in Moscow. 

Some information about tourism in Krasnoyarsk Territories is available at the site: 
$http://tlcom.krs.ru/kalinka/indexe.htm$, \\
tours $http://tlcom.krs.ru/kalinka/indexe.htm$

Every day there are flights from Moscow to Krasnoyarsk airport. Big comfortable airbus IL86 in 4.5 
hours time brings you from Moscow to Krasnoyarsk with good service of KrasAir company and a special 
minivan in 2 hours carries you from Krasnoyarsk airport Yemelianovo directly to the center of 
Zheleznogorsk.

MINING \& CHEMICAL COMBINE has two own rest homes; one of them is in the town territory near the 
forest and another outside of the town not far from it on the bank of Yenisei River. Both of them 
have conference halls, comfortable living rooms and dining rooms. 

\end{document}